\documentclass[prl,12pt,showpacs,amsmath,onecolumn,floatfix]{revtex4}
\usepackage{amsmath}
\usepackage{graphicx}

\newcommand{\ket}[1]{|#1\rangle}
\newcommand{\bra}[1]{\langle#1|}
\newcommand{\avg}[1]{\langle#1\rangle}
\newcommand{\Avg}[1]{\left\langle#1\right\rangle}

\begin{document}
\title{Fundamental Quantum Limit to Multiphoton Absorption Rate
for Monochromatic Light}
\author{Mankei Tsang}
\email{mankei@mit.edu}
\affiliation{Research Laboratory of Electronics,
Massachusetts Institute of Technology,
Cambridge, Massachusetts 02139, USA}

\date{\today}

\begin{abstract}
  The local multiphoton absorption rate for an arbitrary quantum state
  of monochromatic light, taking into account the photon number,
  momentum, and polarization degrees of freedom, is shown to have an
  upper bound that can be reached by coherent fields. This surprising
  result rules out any quantum enhancement of the multiphoton
  absorption rate by momentum entanglement.
\end{abstract}
\pacs{42.50.Ar, 42.50.Dv, 42.50.St}

\maketitle

The excitation of a sample by multiple photons has become an important
tool in optical imaging \cite{lin}. Two-photon absorption was first
predicted by G\"oppert-Mayer in 1931 \cite{mayer}, but because an
intense optical source would be required to observe the phenomenon, it
was only 30 years later, after the invention of lasers, that
two-photon absorption was first experimentally demonstrated by Kaiser
and Garrett \cite{kaiser}.  The advent of mode-locked lasers has
further increased the available optical intensity and contributed to
the success of multiphoton microscopy in biological imaging
\cite{lin,denk}. The spatial resolution improvement by multiphoton
absorption also allows a higher bit density to be recorded in optical
data storage \cite{swainson} and finer features to be written in
lithography \cite{wu,boto}, the primary tool in integrated circuit and
nanostructure fabrication. A particularly intriguing proposal of
``quantum lithography'' was put forth by Boto \textit{et al.}, who
suggest that $N$-photon absorption of $N$ entangled photons can lead
to an $N$-fold resolution enhancement over the Rayleigh-Abbe
resolution limit \cite{boto}. A proof-of-concept experiment was
performed by D'Angelo, Chekhova, and Shih \cite{dangelo}, but current
technology has not yet been able to produce the high flux of entangled
photons required for practical applications.

The requirement of high intensity has motivated researchers to explore
other methods to enhance the multiphoton absorption rate. In quantum
optics, it has long been realized that the statistics of light can
significantly affect the rate of multiphoton processes. For example,
the $N$-photon absorption rate for thermal light is a factor of $N!$
higher than that for laser light with the same intensity
\cite{mollow}, the two-photon absorption rate for weak squeezed light
is proportional to the intensity instead of the intensity squared
\cite{gea}, and the multiphoton absorption rate of spectrally
entangled photons can also depend on the intensity linearly
\cite{javanainen}. These encouraging results have led Boto \textit{et
  al.}\ to suggest that in addition to the resolution enhancement, the
multiphoton absorption rate for $N$ momentum-entangled photons can be
enhanced as well and grow linearly with respect to the intensity, as
the entanglement might constrain the photons to ``arrive at the same
place \cite{boto}.'' Unfortunately, it was later shown that the
quantum effects in the spatial domain are different from those in the
time domain \cite{steuernagel,tsang}. Steuernagel first pointed out
the problem with Boto \textit{et al.}'s claim and studied the
multiphoton absorption rate for four momentum-entangled photons
\cite{steuernagel}, while Tsang showed that there is in general a
trade-off between resolution enhancement and multiphoton absorption
rate for quantum lithography \cite{tsang}.

In Ref.~\cite{tsang}, Tsang also derived an upper bound of the peak
$N$-photon absorption rate for $N$ monochromatic, $s$-polarized
photons in one transverse dimension. The upper bound is on the order
of the classical maximum one-photon intensity raised to the power $N$,
indicating that the nonclassical momentum correlation among $N$
photons is unable to significantly enhance the $N$-photon absorption
rate. This specific result is only applicable to the study of quantum
lithography, and it remains an open but fundamental question whether
photons in an arbitrary quantum state of light can really be
constrained to arrive at the same place and enhance the multiphoton
absorption rate.

In this Letter, using an electromagnetic-field quantization formalism
that takes into account the photon number, momentum, and polarization
degrees of freedom, it is shown that there exists a fundamental upper
bound on the local multiphoton absorption rate for monochromatic
light. The bound is reached by coherent fields as defined by Titulaer
and Glauber \cite{titulaer}, which contain only one excited optical
mode and imply independent photons. Given the well known rate
enhancement effect by photons entangled in the spectral domain
\cite{javanainen}, this result is surprising, as it rules out any
similar effect by momentum entanglement.  The result set forth in this
Letter thus sheds light on our understanding of the quantum nature of
light, and has important implications for the use of quantum optics in
multiphoton imaging applications.

\begin{figure}[htbp]
\centerline{\includegraphics[width=0.3\textwidth]{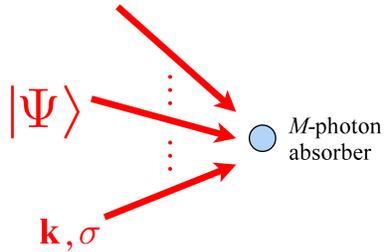}}
\caption{(Color online). An infinitesimally small $M$-photon absorber
  is illuminated by an arbitrary quantum state of monochromatic light,
  given by $\ket{\Psi}$. }
\label{toy}
\end{figure}

Consider the problem of multiphoton absorption depicted in Fig.~1.  An
$M$-photon absorber, such as an atom, a molecule, a quantum dot, or a
nanoparticle, is illuminated by light, with a certain quantum state
$\ket{\Psi}$, in free space. Since such an absorber is usually much
smaller than the characteristic length scales of light, it can be
assumed to be infinitesimally small and interact with the local field
at a certain point in space. The light is assumed to be approximately
monochromatic, which enables one to study the spatial quantum effects
separate from the temporal effects studied in
Refs.~\cite{javanainen}. In most multiphoton absorption experiments,
the absorber is weakly coupled to the electric field of light, so that
the electromagnetic fields can be quantized using the free-space
formalism, as in conventional quantum-optical detection theory
\cite{mandel,glauber}. 

In this case, the positive-frequency electric field
operator is given by \cite{mandel}
\begin{align}
\hat{\boldsymbol{E}}^{(+)}(\boldsymbol{r}t) &=
\frac{i}{(2\pi)^{3/2}}
\sum_{\sigma} \int dk_xdk_y dk_z
\left(\frac{\hbar\omega}{2\epsilon_0}\right)^{1/2}
\nonumber\\&\quad\times
\hat{a}(\boldsymbol{k}\sigma)
\boldsymbol{\varepsilon}(\boldsymbol{k}\sigma)
e^{i\boldsymbol{k}\cdot\boldsymbol{r}-i\omega t},
\end{align}
where $\boldsymbol{k} \equiv k_x\hat{\boldsymbol{x}}
+k_y\hat{\boldsymbol{y}}+k_z\hat{\boldsymbol{z}}$ is the wave vector,
$\omega = ck = c(k_x^2+k_y^2+k_z^2)^{1/2}$ is the optical frequency,
$\sigma$ denotes the two polarizations transverse to the wave
vector, $\boldsymbol{\varepsilon}$ is the unit polarization vector,
and $\hat{a}(\boldsymbol{k}\sigma)$ is the photon annihilation operator
with the commutation relation
$[\hat{a}(\boldsymbol{k}\sigma),\hat{a}^\dagger(\boldsymbol{k}'\sigma')]=
\delta^3(\boldsymbol{k}-\boldsymbol{k}')\delta_{\sigma\sigma'}$.

To conform with classical optics conventions, it is desirable to
change the independent optical mode variables from $(k_xk_yk_z\sigma)$
to $(k_xk_y\omega\gamma\sigma)$, where $\gamma = -1,1$ indicates the
sign of $k_z$ and accounts for both forward and backward propagating
modes, so that $k_z = \gamma(k^2-k_x^2-k_y^2)^{1/2}$. The electric
field can then be expressed in terms of propagation modes
\cite{yuen,tsang}:
\begin{align}
\hat{\boldsymbol{E}}^{(+)}(\boldsymbol{r}t)
&=\frac{i}{(2\pi)^{3/2}}
\sum_{\gamma\sigma}  \int_0^\infty d\omega \int_{k_x^2+k_y^2\le k^2} dk_x dk_y
\nonumber\\&\quad\times
\left(\frac{\hbar\omega}{2\epsilon_0}\right)^{1/2}
\left(\frac{\omega}{c^2|k_z|}\right)^{1/2}
\nonumber\\&\quad\times
\hat{a}(k_xk_y\omega\gamma\sigma)
\boldsymbol{\varepsilon}(k_xk_y\omega\gamma\sigma)
e^{i\boldsymbol{k}\cdot\boldsymbol{r}-i\omega t}.
\end{align}
To make the monochromatic approximation, we follow Blow \textit{et
  al.}\ \cite{blow} and perform the substitutions $\int_0^\infty
d\omega \to 2\pi/T$, $\hat{a}(k_x k_y\omega\gamma\sigma) \to
\hat{a}(k_xk_y\gamma\sigma) [T/(2\pi)]^{1/2}$, and
$\hat{\boldsymbol{E}}^{(+)}(\boldsymbol{r}t) =
\hat{\boldsymbol{E}}^{(+)}(\boldsymbol{r})e^{-i\omega t}$, where $T$
is the characteristic pulse width, resulting in
\begin{align}
\hat{\boldsymbol{E}}^{(+)}(\boldsymbol{r})
&=\frac{i}{(2\pi)^{2}}
\left(\frac{\hbar\omega}{2\epsilon_0c T}\right)^{1/2}
\sum_{\gamma\sigma}  
\int_{k_x^2+k_y^2\le k^2} dk_x dk_y
\nonumber\\&\quad\times
\left(\frac{k}{|k_z|}\right)^{1/2}
\hat{a}(k_xk_y\gamma\sigma)\boldsymbol{\varepsilon}(k_xk_y\gamma\sigma)
e^{i\boldsymbol{k}\cdot\boldsymbol{r}}.
\label{transverse}
\end{align}
In Eq.~(\ref{transverse}), the integration in the transverse momentum
space is inside a circle, which calls for the use of cylindrical
coordinates as defined by $k_x = k\alpha\cos\beta$, $k_y =
k\alpha\sin\beta$, $kx = \rho\cos\phi$, $ky = \rho\sin\phi$, and $kz =
\zeta$.  The electric field becomes
\begin{align}
\hat{\boldsymbol{E}}^{(+)}(\rho\phi\zeta)&=
\frac{i}{2\pi}\left(\frac{\eta_0I_0}{2}\right)^{1/2}
\sum_{\gamma\sigma}  
\int_0^1 d\alpha\int_0^{2\pi}d\beta 
\nonumber\\&\quad\times
\left(\frac{\alpha^2}{1-\alpha^2}\right)^{1/4}
\hat{a}(\alpha\beta\gamma\sigma)
\boldsymbol{\varepsilon}(\alpha\beta\gamma\sigma)
\nonumber\\&\quad\times
e^{i\alpha\rho\cos(\beta-\phi)+
i\gamma(1-\alpha^2)^{1/2}\zeta},
\label{efield}
\end{align}
where $\eta_0 \equiv (\mu_0/\epsilon_0)^{1/2}$ is the free-space
impedance and $I_0$ is defined as $I_0 \equiv
\hbar\omega/(T\lambda^2)$, which is on the order of the optical
intensity of one photon with pulse width $T$ focused onto an area of
$\lambda^2$.  The annihilation operator satisfies the commutation
relation $[\hat{a}(\alpha\beta\gamma\sigma),
\hat{a}^\dagger(\alpha'\beta'\gamma'\sigma')]
=\delta(\alpha-\alpha')\delta(\beta-\beta')
\delta_{\gamma\gamma'}\delta_{\sigma\sigma'}$, and the $s$ and $p$
polarization vectors are $\boldsymbol{\varepsilon}(\alpha\beta\gamma
s) = -\sin(\beta-\phi)\hat{\boldsymbol{\rho}}
+\cos(\beta-\phi)\hat{\boldsymbol{\phi}}$ and
$\boldsymbol{\varepsilon}(\alpha\beta\gamma p) =
-\gamma(1-\alpha^2)^{1/2} [\cos(\beta-\phi)\hat{\boldsymbol{\rho}}
+\sin(\beta-\phi)\hat{\boldsymbol{\phi}}]
+\alpha\hat{\boldsymbol{\zeta}}$, respectively.

An $N$-photon momentum eigenstate can be written as \cite{mandel}
\begin{align}
&\quad \ket{\alpha_1\beta_1\gamma_1\sigma_1,\dots,
\alpha_N\beta_N\gamma_N\sigma_N}
\nonumber\\
&= \frac{1}{\sqrt{N!}}
\hat{a}^\dagger(\alpha_1\beta_1\gamma_1\sigma_1)\dots
\hat{a}^\dagger(\alpha_N\beta_N\gamma_N\sigma_N)\ket{0},
\end{align}
so that a Fock state $\ket{N}$ with a total photon number $N$ can be
expressed in terms of a momentum-space probability amplitude $\phi_N$:
\begin{align}
&\quad\phi_N(\alpha_1\beta_1\gamma_1\sigma_1,\dots,
\alpha_N\beta_N\gamma_N\sigma_N)
\nonumber\\
&\equiv 
\bra{\alpha_1\beta_1\gamma_1\sigma_1,\dots,
\alpha_N\beta_N\gamma_N\sigma_N}N\rangle,
\\
\ket{N} &= \sum_{\gamma_1\sigma_1\dots\gamma_N\sigma_N}
\int d\alpha_1 d\beta_1\dots d\alpha_N d\beta_N
\nonumber\\&\quad\times
\phi_N(\alpha_1\beta_1\gamma_1\sigma_1,\dots,
\alpha_N\beta_N\gamma_N\sigma_N)
\nonumber\\&\quad\times
\ket{\alpha_1\beta_1\gamma_1\sigma_1,\dots,
\alpha_N\beta_N\gamma_N\sigma_N}.
\label{fock}
\end{align}
In the specific case of $N = 2$, $\phi_2$ becomes the well-known
biphoton amplitude, which has been widely used to describe the
entanglement of photon pairs generated by spontaneous parametric
down-conversion \cite{rubin}. A general quantum state of light is then
given by a superposition of Fock states:
\begin{align}
\ket{\Psi} &= \sum_{N=0}^\infty C_N \ket{N},
\end{align}
which completes our description of the quantum state of monochromatic
light in free space.

For $\phi_N$ to be a representation of the $N$-photon quantum state,
$\phi_N$ must satisfy the normalization condition:
\begin{align}
&\sum_{\gamma_1\sigma_1\dots \gamma_N\sigma_N} \int d\alpha_1d\beta_1\dots d\alpha_Nd\beta_N
\nonumber\\&\times
|\phi_N(\alpha_1\beta_1\gamma_1\sigma_1,\dots,\alpha_N\beta_N\gamma_N\sigma_N)|^2 = 1,
\label{norm}
\end{align}
and the bosonic symmetrization condition:
\begin{align}
&\quad\phi_N(\dots,\alpha_n\beta_n\gamma_n\sigma_n,\dots,
\alpha_m\beta_m\gamma_m\sigma_m,\dots)
\nonumber\\
&=\phi_N(\dots,\alpha_m\beta_m\gamma_m\sigma_m,\dots,
\alpha_n\beta_n\gamma_n\sigma_n,\dots)
\nonumber\\
&\quad \textrm{ for any } n \textrm{ and } m.
\end{align}
In particular, a coherent field is defined as a quantum state of light
in which $\phi_N$ is factorizable for all $N$  \cite{titulaer}:
\begin{align}
\phi_N(\alpha_1\beta_1\gamma_1\sigma_1,\dots,
\alpha_N\beta_N\gamma_N\sigma_N) 
= \prod_{n=1}^Nf(\alpha_n\beta_n\gamma_n\sigma_n).
\end{align}
A coherent field is created by exciting only one optical mode, and
implies photons with independent statistics.

The $M$-photon absorption rate for an infinitesimally small absorber
situated at $(\rho,\phi,\zeta)$ in the weak coupling regime is
proportional to
\begin{align}
&\quad\Avg{:\hat{I}_p^M(\rho\phi\zeta):}
\nonumber\\
&=\Avg{:\left\{\frac{1}{\eta_0}
\left[\boldsymbol{p}^*\cdot\hat{\boldsymbol{E}}^{(-)}(\rho\phi\zeta)
\right]
\left[\boldsymbol{p}\cdot\hat{\boldsymbol{E}}^{(+)}(\rho\phi\zeta)
\right]\right\}^M:},
\label{coincidence}
\end{align}
where $\hat{I}_p$ is the optical intensity operator for the electric
field measured along a certain direction with unit vector
$\boldsymbol{p}$. Equation (\ref{coincidence}) also gives the spatial
pattern produced by many independent $M$-photon absorbers.  For a
coherent field, the $M$-photon absorption pattern
$\avg{:\hat{I}_p^M(\rho\phi\zeta):}$ is factorizable and proportional
to $\avg{\hat{I}_p(\rho\phi\zeta)}^M$, and thus agrees with the
classical prediction of the multiphoton absorption pattern
\cite{titulaer}. As such, we can define entanglement for $N$ photons
as a condition in which $\phi_N$ is not factorizable, so that the
multiphoton absorption pattern deviates from the classical theory, as
in the case of quantum lithography \cite{boto}.

As the Fock states are eigenstates of the multiphoton absorption
operator,
we can study the absorption rate for each Fock state and take the
average of the rates at the end of the analysis. Without loss of
generality, assume that the absorber is at the origin. Using
Eqs.~(\ref{efield}) and (\ref{fock}), one can write the $M$-photon
absorption rate for an $N$-photon state explicitly in terms of
$\phi_N$:
\begin{widetext}
\begin{align}
\bra{N}:\hat{I}_p^M:\ket{N}
&=\left(\frac{I_0}{8\pi^2}\right)^{M} \frac{N!}{(N-M)!}
\sum_{\gamma_{M+1}\sigma_{M+1}\dots \gamma_N\sigma_N}\int d\alpha_{M+1}d\beta_{M+1}
\dots d\alpha_{N}d\beta_{N}
\nonumber\\&\quad\times
\Bigg|
\sum_{\gamma_1\sigma_1\dots \gamma_M\sigma_M}
\int d\alpha_1d\beta_1 \dots d\alpha_M d\beta_M
\left[\prod_{n=1}^M \left(\frac{\alpha_n^2}{1-\alpha_n^2}\right)^{1/4}
\boldsymbol{p}\cdot
\boldsymbol{\varepsilon}(\alpha_n\beta_n\gamma_n\sigma_n)
\right]
\nonumber\\&\quad\times
\phi_N(\alpha_1\beta_1\gamma_1\sigma_1,\dots,
\alpha_N\beta_N\gamma_N\sigma_N)
\Bigg|^2.
\label{inner}
\end{align}
To derive an upper bound on this quantity, we observe that the
$M$-dimensional integral in Eq.~(\ref{inner}) can be regarded as an
inner product between the expression in square brackets and
$\phi_N^*$. Applying Schwarz's inequality and the normalization
condition in Eq.~(\ref{norm}), we obtain
\begin{align}
\bra{N}:\hat{I}_p^M:\ket{N}
&\le \left(\frac{I_0}{8\pi^2}\right)^{M} \frac{N!}{(N-M)!}
\sum_{\gamma_{M+1}\sigma_{M+1}\dots \gamma_N\sigma_N}\int d\alpha_{M+1}d\beta_{M+1}
\dots d\alpha_{N}d\beta_{N}
\nonumber\\&\quad\times
\sum_{\gamma_1\sigma_1 \dots \gamma_M\sigma_M}
\int d\alpha_1d\beta_1 \dots d\alpha_M d\beta_M
\left|\phi_N(\alpha_1\beta_1\gamma_1\sigma_1,\dots,
\alpha_N\beta_N\gamma_N\sigma_N)
\right|^2
\nonumber\\&\quad\times
\sum_{\gamma_1\sigma_1\dots \gamma_M\sigma_M}
\int d\alpha_1d\beta_1 \dots d\alpha_M d\beta_M
\left|\prod_{n=1}^M \left(\frac{\alpha_n^2}{1-\alpha_n^2}\right)^{1/4}
\boldsymbol{p}\cdot\boldsymbol{\varepsilon}(\alpha_n\beta_n\gamma_n\sigma_n)
\right|^2
\nonumber\\
&= \left(\frac{I_0}{8\pi^2}\right)^{M} \frac{N!}{(N-M)!}
\left[
\sum_{\gamma\sigma}\int_0^1 d\alpha \int_0^{2\pi}d\beta
\left(\frac{\alpha^2}{1-\alpha^2}\right)^{1/2}
\left|\boldsymbol{p}\cdot
\boldsymbol{\varepsilon}(\alpha\beta\gamma\sigma)\right|^2\right]^M
\nonumber\\&
=\left(\frac{I_0}{3\pi}\right)^M \frac{N!}{(N-M)!}.
\label{bound}
\end{align}
\end{widetext}
This bound does not depend on $\boldsymbol{p}$, the direction along
which the electric field is measured, and is therefore applicable to
the isotropic multiphoton absorption measurement $\avg{:\hat{I}^M:} =
\avg{:[\hat{\boldsymbol{E}}^{(-)}\cdot\hat{\boldsymbol{E}}^{(+)}]^M:}$.
Hence, for an arbitrary quantum state, the $M$-photon absorption rate
is bounded by the following:
\begin{align}
\Avg{:\hat{I}^M:}
\le \left(\frac{I_0}{3\pi}\right)^M 
\sum_{N\ge M} |C_N|^2\frac{N!}{(N-M)!}.
\label{central}
\end{align}
Equation (\ref{central}) is the central result of this Letter.
Although this bound is derived for one absorber, it is also equivalent
to a bound on the peak absorption rate for many independent
absorbers. The factor $\sum_N |C_N|^2 N!/(N-M)!$ depends only on the
statistics of total photon number and accounts for the effect of
photon-number fluctuations on the multiphoton absorption rate, as
investigated in Refs.~\cite{mollow,gea}. This factor, however, does
not depend on the structure of $\phi_N$, which governs the momentum
correlations of photons. The dependence of the bound on the $M$th
power of $I_0$, on the other hand, agrees with classical multiphoton
absorption theory, and suggests that a coherent field can reach this
upper bound. To prove this, recall the fact that the Schwarz upper
bound is reached when the two functions in the inner product are
linearly dependent. One possible quantum state with a $\phi_N^*$
linearly dependent on the square-bracketed expression in
Eq.~(\ref{inner}) is given by
\begin{align}
\phi_N \propto \prod_{n=1}^N
\left(\frac{\alpha_n^2}{1-\alpha_n^2}\right)^{1/4}
\boldsymbol{p}^*\cdot
\boldsymbol{\varepsilon}^*(\alpha_n\beta_n\gamma_n\sigma_n),
\label{max_coherent}
\end{align}
which is factorizable and thus a coherent field by definition.

In conclusion, the quantum limit to multiphoton absorption rate
derived above demonstrates that, heuristically speaking, it is
impossible for monochromatic photons to arrive at the same specific
location in free space more often than do independent photons focused
onto an area of $\lambda^2$.  While this result is applicable to most
multiphoton imaging experiments and suggests that spatial quantum
effects are not useful in enhancing the multiphoton absorption rate
for those applications, it is possible to generalize the theory to
more complex and exotic situations, such as the use of polychromatic
light, cavity confinement, and strong coupling between light and the
absorbers. Such generalizations will lead to the study of multimode
cavity quantum electrodynamics, which should exhibit more complex and
interesting phenomena than the case studied here and may ultimately
benefit nonlinear optics applications.



The author would like to acknowledge helpful discussions with Jeffrey
Shapiro and Seth Lloyd. This work is supported by the W.\ M.\ Keck
Foundation Center for Extreme Quantum Information Theory.

\end{document}